\documentclass{webofc}
\usepackage[varg]{txfonts}   
\begin{document}

\def\a{\alpha}
\def\b{\beta}
\def\d{{\delta}}
\def\l{\lambda}
\def\e{\epsilon}
\def\p{\partial}
\def\m{\mu}
\def\n{\nu}
\def\t{\tau}
\def\th{\theta}
\def\s{\sigma}
\def\g{\gamma}
\def\o{\omega}
\def\r{\rho}
\def\D{\Delta}
\def\G{{\Gamma}}
\def\half{\frac{1}{2}}
\def\hatt{{\hat t}}
\def\hatx{{\hat x}}
\def\hatp{{\hat p}}
\def\hatX{{\hat X}}
\def\hatY{{\hat Y}}
\def\hatP{{\hat P}}
\def\haty{{\hat y}}
\def\z{\zeta}
\def\whatX{{\widehat{X}}}
\def\whata{{\widehat{\alpha}}}
\def\whatb{{\widehat{\beta}}}
\def\whatV{{\widehat{V}}}
\def\hatth{{\hat \theta}}
\def\hatta{{\hat \tau}}
\def\hatrh{{\hat \rho}}
\def\hatva{{\hat \varphi}}
\def\barx{{\bar x}}
\def\bary{{\bar y}}
\def\barz{{\bar z}}
\def\baro{{\bar \omega}}
\def\barpsi{{\bar \psi}}
\def\sp{\sigma^\prime}
\def\nn{\nonumber}
\def\cb{{\cal B}}
\def\2pap{2\pi\alpha^\prime}
\def\wideA{\widehat{A}}
\def\wideF{\widehat{F}}
\def\beq{\begin{eqnarray}}
 \def\eeq{\end{eqnarray}}
 \def\4pap{4\pi\a^\prime}
 \def\xp{{x^\prime}}
 \def\sp{{\s^\prime}}
 \def\spp{{\s^{\prime\prime}}}
 \def\ap{{\a^\prime}}
 \def\tp{{\t^\prime}}
 \def\zp{{z^\prime}}
 \def\op{\omega^\prime}
 \def\xpp{x^{\prime\prime}}
 \def\xppp{x^{\prime\prime\prime}}
 \def\barxp{{\bar x}^\prime}
 \def\barxpp{{\bar x}^{\prime\prime}}
 \def\barxppp{{\bar x}^{\prime\prime\prime}}
 \def\barchi{{\bar \chi}}
 \def\baro{{\bar \omega}}
 \def\bpsi{{\bar \psi}}
 \def\barg{{\bar g}}
 \def\barz{{\bar z}}
 \def\bareta{{\bar \eta}}
 \def\ta{{\tilde \a}}
 \def\tb{{\tilde \b}}
 \def\tc{{\tilde c}}
 \def\tz{{\tilde z}}
 \def\tJ{{\tilde J}}
 \def\tpsi{\tilde{\psi}}
 \def\tal{{\tilde \alpha}}
 \def\tbe{{\tilde \beta}}
 \def\tga{{\tilde \gamma}}
 \def\tchi{{\tilde{\chi}}}
 \def\barth{{\bar \theta}}
 \def\bareta{{\bar \eta}}
 \def\barom{{\bar \omega}}
 \def\bole{{\boldsymbol \epsilon}}
 \def\bolth{{\boldsymbol \theta}}
 \def\bomega{{\boldsymbol \omega}}
 \def\bolmu{{\boldsymbol \mu}}
 \def\bola{{\boldsymbol \alpha}}
 \def\bolb{{\boldsymbol \beta}}
 \def\bolX{{\boldsymbol X}}
 \def\boln{{\boldsymbol n}}
 \def\bba{{\boldsymbol a}}
 \def\bbk{{\boldsymbol k}}
 \def\bbA{{\boldsymbol A}}
 \def\bbP{{\boldsymbol P}}
 \def\bbp{{\boldsymbol p}}
 \def\mathP{{\mathbb P}}

%
\title{Gravitational Scattering Amplitudes and \\Closed String Field Theory 
in the Proper-Time Gauge}
%
%

\author{\firstname{Taejin} \lastname{Lee}\inst{1}
\thanks{\email{taejin@kangwon.ac.kr}} 
}

\institute{Department of Physics, Kangwon National University, Chuncheon 24341
Korea
          }

\abstract{
We construct a covariant closed string field theory by extending recent works on the covariant open string field 
theory in the proper-time gauge. Rewriting the string scattering amplitudes generated by the closed string field 
theory in terms of the Polyakov string path integrals, we identify the Fock space representations of the closed string vertices. We show that the Fock space representations of the closed string field theory may be completely factorized into those of the open string field theory. It implies that the well known Kawai-Lewellen-Tye (KLT) relations of the first quantized string theory may be promoted to the second quantized closed string theory. 
We explicitly calculate the scattering amplitudes of three gravitons by using the closed string field theory in the proper-time gauge. 
}

\maketitle
\section{Introduction}
\label{intro}
Construction of a finite quantum theory of gravity was the main motivation for theorists to turn towards 
the string theory \cite{Scherk74} as it has become clear that the perturbative theory of gravity based on the quantum field theory \cite{Dewitt1, Dewitt2, Dewitt3} is not renormalizable \cite{tHooft74, Veltman76}. Given that the spectrum of a free closed string contains the spin two massless particle state, which corresponds to the graviton and the perturbative string theory is expected to be free from ultraviolet divergences, it stands to reason that the string theory should receive attention from theorists 
as a consistent framework for unifying all quantum theories of fundamental forces, including gravity. 
In view of this, the closed string field theory is anticipated to be a passage to the finite quantum theory of gravity. However, despite years of effort, construction of a consistent perturbative closed string field theory has not been yet completed. 

In this work, we construct the covariant closed string field theory by extending recent works on open string field theory in the proper-time gauge \cite{TLee88ann, TLee2017plb, TLee1609, TLeemulti2017, Lai2017scatt}. Following the general study on the classical limit of quantum gravity \cite{Boulware75}, 
the closed string field theory as a consistent quantum theory of gravity should reduce to the 
classical Einstein gravity in the low energy domain. For a comparison of the proposed closed 
string field theory with the Einstein gravity, we will calculate the gravitational scattering amplitudes at tree level and show that the closed string field theory correctly reproduces that of the conventional Einstein 
gravity. In the process of evaluating the string scattering, we also find that the Fock space representations 
of the closed string field theory are completely factorized into those of the open string field theory in general: 
This factorization indicates that the KLT \cite{Kawai86} relations of the first quantized string theory may be fully extended to the second quantized string theory.

\section{Closed string field theory in the proper-time gauge}
\label{sec closed}
\begin{figure}[h]
\centering
\sidecaption
\includegraphics[width=3cm,clip]{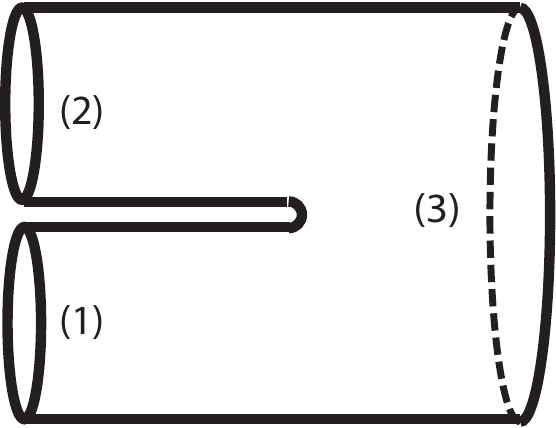}
\caption{Closed string interaction in the proper-time gauge.}
\label{stringinteraction}      
\end{figure}

The closed string field theory in the proper-time may be formally described by the action 
\beq
S &=& \langle \Phi \vert {\cal K} \Phi \rangle + \frac{g}{3} \Biggl(\langle \Phi \vert \Phi \circ \Phi \rangle + \langle \Phi \circ \Phi \vert \Phi \rangle \Biggr) \nn
\eeq
where ${\cal K} = L_0 - i\e$ and the second term denotes the closed string interaction as depicted in 
Fig. \ref{stringinteraction}
In the proper-time gauge we fix the length parameters of three strings as $\a_1 = \a_2 =1$ and 
$\a_3 =-2$. Because the length parameters can be written in terms of the two dimensional metric on the string world-sheet, fixing the length parameters is equivalent to fixing the metric on the world-sheet by using the reparametrization invariance. The kinetic term is indpendent of the length parameters if we adopt the Fock space representation for the the string field \cite{TLee1609}. Although the string field action, describing the off-shell dynamics, may depend on the length parameters, the physical scattering amplitudes, which are generated by the string field action, may be independent of the length parameters. 

\section{Fock space representation of the closed string field theory}

Construction of the Fock space representation of the closed string field theory begins with the Schwarz-Christoffel
(SC) mapping from the string world-sheet to the complex plane \cite{TLee1609}: 
\beq
\rho = \ln (z-1) + \ln z .  
\eeq 
The local coordinates $\zeta_r = \xi_r + i \eta_r$, $r=1, 2, 3$ 
defined on invidual string world sheet patches are related to $z$ as follows, 
\begin{subequations}
\beq
e^{-\zeta_1} &=& e^{\t_0} \frac{1}{z(z-1)}, \\
e^{-\zeta_2} &=& - e^{\t_0} \frac{1}{z(z-1)}, \\
e^{-\zeta_3} &=& - e^{- \frac{\t_0}{2}} \sqrt{z(z-1)} .
\eeq
\end{subequations}    
Then, by using the SC mapping, we may write the the Green's function on the world sheet as follows:  
\beq \label{greenclosed}
G_C(\r_r, \rho^\prime_s) &=& \ln \vert z_r - \zp_s \vert \nn\\
&=& - \d_{rs} \left\{\sum_{n=1} \frac{e^{-n\D}}{2n} \left(e^{in(\eta^\prime_s- \eta_r)} + e^{-in(\eta^\prime_s- \eta_r)} \right) - \max(\xi, \xi^\prime)  \right\} \nn\\
&& + \sum_{n, m} \bar C^{rs}_{nm} e^{|n| \xi_r + |m| \xi^\prime_s} e^{in\eta_r} e^{im\eta^\prime_s},
\eeq
where $\bar C^{rs}_{nm}$ is the Fourier components of the Green's function on the string world sheet. 

The closed string field theory action generates perturbative string Feynmann diagrams corresponding to the 
multi-string scattering amplitudes
\beq
{\cal W}_N 
&=&\langle  {\bf P} \vert \exp\left( \sum_r \xi_r L^{(r)}_0 \right) \vert V[N] \rangle .
\eeq 
For three-string scattering ${\cal W}_3$, by employing the oscillatory basis, we obtain the Fock space representation of the three-string-vertex $\vert V[3]$:
\beq
\vert V[3] \rangle &=&  \exp  \Biggl\{
\sum_{r,s} \Bigl( \sum_{n, m \ge 1} \frac{1}{2} \bar N^{rs}_{nm}\,\frac{\a^{(r)\dag}_n}{2} \cdot 
\frac{\a^{(r)\dag}_m}{2} 
+ \sum_{n \ge 1}\bar N^{rs}_{n0} \frac{\a^{(r)\dag}_n}{2} \cdot \frac{p^{(s)}}{2} \Bigr) 
\Biggr\} \nn\\
&&  \exp \Biggl\{ \t_0 \sum_r \frac{1}{\a_r} \left(\half \left(\frac{p^{(r)}}{2}\right)^2 -1 \right) 
\Biggr\}  \nn\\
&& \exp  \Biggl\{
\sum_{r,s} \Bigl( \sum_{n, m \ge 1} \frac{1}{2} \bar N^{rs}_{nm}\,\frac{\tilde\a^{(r)\dag}_n}{2} \cdot \frac{\tilde\a^{(r)\dag}_m}{2} 
+ \sum_{n \ge 1}\bar N^{rs}_{n0} \frac{\tilde\a^{(r)\dag}_n}{2} \cdot \frac{p^{(s)}}{2} \Bigr) \Biggr\}\nn\\
&& \exp \Biggl\{ \t_0 \sum_r \frac{1}{\a_r} \left(\half \left(\frac{p^{(r)}}{2}\right)^2 -1 \right) 
\Biggr\} \vert 0 \rangle, \label{3vertex}
\eeq  
where $\bar N^{rs}_{nm}$, $r, s = 1, 2, 3$ are the Neumann functions of the open string theory. 
From the explicit expression of the Fock space representation of the three-closed-string vertex, it is 
evident that the closed string vertex may be completely factorized into those of open string vertices. This 
observation suggests that the KLT relations may be fully extended to the second quantized string theory. 

\section{Three-graviton scattering amplitude}

Having obtained the Fock space representation of the three-closed-string vertex, we shall calculate the three-graviton-scattering amplitude to confirm that the closed string field theory, as a consistent quantum theory 
of gravity reduces to the Einstein gravity in the low energy domain. We may decompose the masless closed string 
states $h_{\m\n} \,\a^{\m}_{-1} \tilde \a^{\n}_{-1}\vert 0 \rangle$ into the graviton, the anti-symmetric ternsor, and the scalar field states:
\beq
h_{\m\n}  &=&\Bigl\{ \half \left(h_{\m\n} + h_{\n\m} \right) - \eta_{\m\n} \frac{d}{d-2} h^\s{}_\s \Bigr\}
+ \Bigl\{\half \left(h_{\m\n} - h_{\n\m} \right) \Bigr\} + \eta_{\m\n} \Bigl\{\frac{1}{d-2} h^\s{}_\s \Bigr\}.
\eeq 
For the three-graviton scattering, we choose the symmetric traceless part and write the external string state
as 
\beq
\vert \Psi_{3G} \rangle = \prod_{r=1}^3 \left\{h_{\m\n}(p^r) \a^{(r)\m}_{-1} \tilde \a^{(r)\n}_{-1} \right\}\vert 0 \rangle . \label{3G}
\eeq  
The three-graviton scattering amplitude follows from the explicit expression of the three-string vertex Eq.(\ref{3vertex}) with the three-graviton state Eq. (\ref{3G}): 
\beq
{\cal A}_{3G} &=& \int \prod_{r=1}^3 dp^{(r)} \d \left(\sum_{r=1}^3 p^{(r)}\right) \frac{2g}{3} \, \langle \Psi_{3G}\vert  V[3] \rangle \nn\\
&=& \left(\frac{2g}{3}\right) e^{-2 \t_0 \sum_{r=1}^3 \frac{1}{\a_r}} \int \prod_{i=1}^3 dp^{(i)} \d\left(
\sum_{i=1}^3 p^{(i)} \right) \nn\\
&& \langle 0 \vert \left\{\prod_{i=1}^3 h_{\m\n}(p^{(i)}) a^{(i)\m}_1 \cdot \tilde a^{(i)\n}_1 \right\} \frac{1}{2^5}
\left(\sum_{r, s =1}^3 \bar N^{rs}_{11} a^{(r)\dag}_1 \cdot a^{(s)\dag}_1 \right) \nn\\
&& \left(\sum_{t=1}^3 \bar N^t_1 a^{(t)\dag}_1 \cdot \bbp \right) \frac{1}{2^5}
\left(\sum_{l, m =1}^3 \bar N^{lm}_{11} \tilde a^{(l)\dag}_1 \cdot \tilde a^{(m)\dag}_1 \right)\left(\sum_{n=1}^3 \bar N^n_1 \tilde a^{(n)\dag}_1 \cdot \bbp \right) \vert 0 \rangle ,
\eeq 
where $\bbp = p^{(1)}-p^{(2)}$.
Making use of the Neumann functions of the open string given in Ref. \cite{TLee1609}, 
\begin{subequations}
\beq
\bar N^{11}_{11} &=& \frac{1}{2^4}, ~~~ \bar N^{22}_{11} = \frac{1}{2^4}, ~~~ \bar N^{33}_{11} = 2^2, \label{neumanna}\\
\bar N^{12}_{11} &=& \bar N^{21}_{11} = \frac{1}{2^4}, ~~~ \bar N^{23}_{11} = \bar N^{32}_{11} = \half, 
~~~ \bar N^{31}_{11} = \bar N^{13}_{11} = \half ,\label{neumannb}\\
\bar N^1_1 &=& \bar N^2_1 = \frac{1}{4}, ~~~ \bar N^3_1 = -1 , \label{neumannc}
\eeq
\end{subequations}
we are able to evaluate 
the three-graviton amplitude 
\beq
{\cal A}_{3G} &=&  \left(\frac{2g}{3}\right) 2^6 \left(\frac{1}{2^5}\right)^2 \int \prod_{i=1}^3 dp^{(i)} \d\left(\sum_{i=1}^3 p^{(i)} \right) h_{\m_{1}\n_{1}}(p^{(1)})h_{\m_{2}\n_{2}}(p^{(2)})h_{\m_{3}\n_{3}}(p^{(3)})\nn\\
&& \Biggl\{
-\frac{1}{2^4} \eta^{\m_1\m_2} \bbp^{\m_3} + \frac{1}{2^3} \eta^{\m_1\m_3} \bbp^{\m_2} + \frac{1}{2^3}\eta^{\m_2\m_3} \bbp^{\m_1} - \frac{1}{2^4}\eta^{\m_2\m_1} \bbp^{\m_3} + 
\frac{1}{2^3} \eta^{\m_3\m_1} \bbp^{\m_2} \nn\\
&&+ \frac{1}{2^3} \eta^{\m_3\m_2} \bbp^{\m_1} \Biggr\}
 \Biggl\{-\frac{1}{2^4}\eta^{\n_1\n_2} \bbp^{\n_3} + \frac{1}{2^3}\eta^{\n_1\n_3} \bbp^{\n_2} + \frac{1}{2^3}\eta^{\n_2\n_3} \bbp^{\n_1}- \frac{1}{2^4}\eta^{\n_2\n_1} \bbp^{\n_3} \nn\\
&& + 
\frac{1}{2^3}\eta^{\n_3\n_1} \bbp^{\n_2} + \frac{1}{2^3} \eta^{\n_3\n_2} \bbp^{\n_1} \Biggr\}.
\eeq
 Repeatedly making use of the covariant gauge condition $\p^\m h_{\m\n} = 0$, 
which is equivalent to the de Donder gauge condition for the graviton sector,
$\p^\m h_{\m\n} - \frac{1}{d-2} \p_\n h^\s{}_\s = 0$, we find 
\beq
{\cal A}_{3G}
&=& \kappa
\int \prod_{i=1}^3 dp^{(i)} \d\left(\sum_{i=1}^3 p^{(i)} \right)  h_{\m_{1}\n_{1}}(p^{(1)})h_{\m_{2}\n_{2}}(p^{(2)})h_{\m_{3}\n_{3}}(p^{(3)}) \nn\\
&& \Bigl\{
\eta^{\m_1\m_2} p^{(1)\m_3} + \eta^{\m_2\m_3} p^{(2)\m_1} 
+ \eta^{\m_3\m_1} p^{(3)\m_2} \Bigr\}\Bigl\{\eta^{\n_1\n_2} p^{(1)\n_3} + \eta^{\n_2\n_3} p^{(2)\n_1} 
+ \eta^{\n_3\n_1} p^{(3)\n_2} \Bigr\}, 
\eeq
where $\kappa =\frac{g}{2^7 \cdot 3}= \sqrt{32 \pi G_{10}}$. ${\cal A}_{3G}$ is precisely the three-graviton 
interaction term of the conventional perturbative Einstein gravity \cite{Sannan86}.

\section{Conclusions}

By extending recent works on the open string field theory in the proper-time gauge \cite{TLee2017plb, TLee1609, TLeemulti2017, Lai2017scatt}, we constructed 
a covariant closed string field theory. The closed string field theory generates perturbative three string 
diagrams, which may be re-expressed in terms of the Polyakov string path-integral. By mapping the three-string
world sheet diagram onto the complex plane, we evaluate explicitly the three-string-scattering amplitude 
and obtain the Fock space representation of the three-string-vertex. For the three-string diagram, we may map the positions of the external strings onto the real line by using the $SL(2,C)$ invariance. Then, it 
follows that the Fourier components of the closed string Green's function on the string world sheet 
may be written in terms of those of the open string, {\it i.e.}, the Neumann functions of the open string.  
We also find that the Fock-space representation of the closed three-string-vertex may be completely factorized 
into those of the open string in the proper-time gauge. This factorization of the closed string vertex may be a string field theoretical extension of the KLT relations, previously studied in the first quantized string theory. To confirm that the 
closed string field theory in the proper-time gauge correctly reduces to the perturbative Einstein gravity, 
we calculate the three-graviton-scattering amplitude. 

More work remains to be done to complete the construction of the covariant closed string field theory in the 
proper-time gauge: 1) Fock space representations of the general closed string amplitudes should be constructed. 
2) We may explicitly evaluate the four-graviton scattering amplitudes. 3) Full extensions of the KLT relations 
at the level of the second quantized string theory would be certainly among the remaining tasks to be done.
We will discuss these subjects in papers sequel to this work.

\vskip 0.5cm

{\bf Acknowledgments}
This work was supported by Basic Science Research Program through the National Research Foundation of Korea (NRF) funded by the Ministry of Education (2017R1D1A1A02017805).

%

\begin{thebibliography}{}
%
%
\bibitem{Scherk74}
J. Scherk and J. H. Schwarz, 
Nucl. Phys. B {\bf81}, 118 (1974).

\bibitem{Dewitt1}
B. S. DeWitt, 
Phys. Rev. {\bf 160}, 1113 (1967).

\bibitem{Dewitt2}
B. S. DeWitt, 
Phys. Rev. {\bf 162}, 1195 (1967). 

\bibitem{Dewitt3}
B. S. DeWitt, 
Phys. Rev. {\bf 162}, 1239 (1967).

\bibitem{tHooft74}
G. ’t Hooft and M. J. Veltman, 
Ann. Inst. Henri Poincare, A {\bf 20}, 69 (1974).

\bibitem{Veltman76}
M. J. Veltman, 
in R. Balian and J. Zinn-Justin, eds., {\it Methods in Field Theory, Proceedings of the Les Houches Summer School 1975}, (North-Holland, Amsterdam, 1976) 265–327. 

\bibitem{TLee88ann}
T. Lee, Ann. Phys. {\bf 183}, 191 (1988).

\bibitem{TLee2017plb}
T. Lee,
Phys. Lett. B {\bf 768}, 248 (2017).

\bibitem{TLee1609}
T. Lee, 
{\it Covariant open bosonic string field theory on multiple D-branes in the proper-time gauge},
to be published in Jour. Kor. Soc. (2017),
arXiv:1609.01473 [hep-th].

\bibitem{TLeemulti2017}
T. Lee, 
arXiv:1703.06402 [hep-th].

\bibitem{Lai2017scatt}
S. H. Lai, J. C. Lee, Yi Yang, and T. Lee,  
arXiv:1706.08025 [hep-th].

\bibitem{Boulware75}
D. G. Boulware and S. Deser, 
Ann. Phys. {\bf 89}, 193 (1975).

\bibitem{Kawai86}
H. Kawai, D. C. Lewellen, and S. H. Tye, 
Nucl. Phys. B {\bf 269}, 1 (1986). 

\bibitem{Sannan86}
S. Sannan, 
Phys. Rev. D {\bf 34}, 1749 (1986).




\end{thebibliography}
%
%

\end{document}